\documentclass[conference]{IEEEtran}
\IEEEoverridecommandlockouts

\usepackage{xcolor,soul,framed} 

\colorlet{shadecolor}{yellow}
\usepackage[pdftex]{graphicx}
\graphicspath{{../pdf/}{../jpeg/}}
\DeclareGraphicsExtensions{.pdf,.jpeg,.png}

\usepackage[cmex10]{amsmath}
\usepackage{array}
\usepackage{mdwmath}
\usepackage{mdwtab}
\usepackage{eqparbox}
\usepackage{url}

\usepackage{amssymb}
\graphicspath{ {./images/} }
\usepackage{orcidlink}
\usepackage{breqn}
\usepackage{physics}

\usepackage{tikz} 

\usepackage{float}

\usepackage{cite}
\usepackage{amsmath,amssymb,amsfonts}
\usepackage{amsmath}
\usepackage{algorithm}
\usepackage{algorithmic}
\usepackage{graphicx}
\usepackage{textcomp}
\usepackage{xcolor}
\def\BibTeX{{\rm B\kern-.05em{\sc i\kern-.025em b}\kern-.08em
    T\kern-.1667em\lower.7ex\hbox{E}\kern-.125emX}}

\usepackage{lipsum}
\usepackage{diagbox}

\begin{document}

\title{A Novel Approach to Classify Power Quality Signals Using Vision Transformers}

\author{
\IEEEauthorblockN{Ahmad Mohammad Saber\orcidlink{0000-0003-3115-2384}}
\IEEEauthorblockA{\textit{Electrical and Computer Engineering} \\
\textit{University of Toronto}\\
Toronto, ON, Canada \\
ahmad.m.saber@ieee.org}
\and
\IEEEauthorblockN{ Alaa Selim\orcidlink{0000-0002-5077-5168}}
\IEEEauthorblockA{\textit{Electrical and Computer Engineering} \\
\textit{University of Connecticut}\\
Storrs, USA \\
alaa.selim@uconn.edu}
\and
\IEEEauthorblockN{Mohamed M. Hammad}
\IEEEauthorblockA{\textit{Renewable Energy Engineering} \\
\textit{Siemens Energy, Egypt }\\
Mohamed.mahmoud-mostafa-\\
hammad@siemens-energy.com}
\and
\IEEEauthorblockN{ }
\IEEEauthorblockA{\textit{ } \\
\textit{ }\\
  \\
 }
\and
\IEEEauthorblockN{\textcolor{white}{..........}Amr Youssef\orcidlink{0000-0002-4284-8646}}
\IEEEauthorblockA{\textit{\textcolor{white}{..........}CIISE} \\
\textit{\textcolor{white}{..........}Concordia University}\\
\textcolor{white}{..........}Montreal, QC, Canada \\
\textcolor{white}{..........}youssef@ciise.concordia.ca}
\and
\IEEEauthorblockN{Deepa Kundur\orcidlink{0000-0001-5999-1847}}
\IEEEauthorblockA{\textit{Electrical and Computer Engineering} \\
\textit{University of Toronto}\\
Toronto, ON, Canada \\
dkundur@ece.utoronto.ca}
\and
\IEEEauthorblockN{Ehab El-Saadany\orcidlink{0000-0003-0172-0686}}
\IEEEauthorblockA{\textit{Electrical Engineering} \\
\textit{Khalifa University}\\
Abu Dhabi, UAE \\
ehab.elsadaany@ku.ac.ae}
}

\maketitle

\begin{abstract}

With the rapid integration of electronically interfaced renewable energy resources and loads into smart grids, there is increasing interest in power quality disturbances (PQD) classification to enhance the security and efficiency of these grids. This paper introduces a new approach to PQD classification based on the Vision Transformer (ViT) model. When a PQD occurs, the proposed approach first converts the power quality signal into an image and then utilizes a pre-trained ViT to accurately determine the class of the PQD. Unlike most previous works, which were limited to a few disturbance classes or small datasets, the proposed method is trained and tested on a large dataset with 17 disturbance classes. Our experimental results show that the proposed ViT-based approach achieves PQD classification precision and recall of 98.28\% and 97.98\%, respectively, outperforming recently proposed techniques applied to the same dataset.

\end{abstract}


\section{Introduction}

With the rise of distributed energy resources and advanced power electronics in today's electric grids, the occurrence of Power Quality Disturbances (PQDs) has become more common. These disturbances, which can cause fluctuations in voltage, frequency, and waveform quality, present a significant challenge for maintaining the stability and efficiency of the grid  \cite{Tian9652053,singh2009power}. As the grid becomes more complex, accurately identifying and managing these disturbances becomes increasingly important for utility companies and grid operators \cite{chawda2020comprehensive}.

Classifying PQDs is crucial because it helps ensure a reliable and efficient power supply. When PQDs are accurately identified, utilities can quickly address issues, preventing potential damage to equipment and optimizing energy usage. This not only helps meet regulatory standards but also improves customer satisfaction by reducing power disruptions. Additionally, having a clear understanding of these disturbances supports better decision-making in grid planning and maintenance, leading to greater energy efficiency and cost savings \cite{saber2024enhancing}. Therefore, there is a great necessity for accurate PQD classification to maintain the security and efficiency of modern electrical systems.

Classically, PQD classification was performed using one or more hand-crafted features.
Meta-heuristic optimization techniques, e.g., particle swarm optimization \cite{ahila2015integrated} and artificial bee colony \cite{khokhar2017new},  were used to select the most important features for PQD classification.
In this approach, the original feature set may still contain many irrelevant features, which consumes a long time during the optimization process \cite{wang2019novel}.

Zhao \textit{et al.} proposed using Stockwell Transform (ST) for signal processing and feature selection and then using Decision Trees for PQD classification \cite{zhao2019power}. An approach for PQD classification using the Independent Component Analysis followed by the Wavelet Transform and then Random Forests was introduced in \cite{liu2020high}. Rodriguez \textit{et al.} \cite{rodriguez2021classification} used the Hilbert-Huang transform (HHT) with the Long Short-Term Memory (LSTM) model and could handle 9 disturbance classes. Furthermore, Ma \textit{et al.} developed a modified ST combined with the Multi-scale Parallel Attention Residual Network model \cite{ma2023complex}.

Convolutional neural networks (CNNs) have gained popularity in various classification tasks in recent years. CNNS can use convolutional kernels to estimate weights for each feature and thus obtain higher performance due to the reduction of the parameter number. A PQD classification method based on combining the ST with CNNs was proposed in \cite{cui2022detection}. Sindi \textit{et al.} proposed using the CNN model for PQD classification in a hybrid way by generating a set of features from the power signals in 1D form and the signal images in 2D form \cite{sindi2021novel}. These feature vectors are then combined and fed to a fully connected layer for classification. The ST was used to obtain the optimum contour image of a PQD signal and then the resulting image files are classified by using the CNN algorithm \cite{eristi2022new}. Ozer \textit{et al.} used the CNN along with the bidirectional LSTM model for the classification of PQDs  \cite{ozer2021cnn}. Finally, Carni and Lamonaca proposed combining the HHT with the CNN model for PQD classification \cite{9832923}. A PQD classification framework based on time series transformers was proposed \cite{saber2024enhancing}.

Despite the previous efforts, most of the existing methods either lacked accuracy or were assessed on datasets with few PQD classes or small sizes. To tackle this gap, this paper presents a new approach for accurate PQD classification using the ViT model. The proposed approach converts a given PQD signal into an image and then leverages a pre-trained ViT model to accurately predict the class of the signal. This approach allows us to exploit the information embedded in the signal as an image, compared to previous approaches that treat the signal as independent numerical values.  
Additionally, to the best of the authors' knowledge, ViTs have not been investigated in the context of power systems yet. To summarize, our main contributions in this paper can be summarized as follows:

\begin{enumerate}
    \item A novel framework for PQDs classification using the ViT model, which has not been used before in any of the related works. 
    \item The assessment of the proposed framework is performed on a large dataset with more than 250,000 samples and 17 classes of PQDs including single and multiple simultaneous events.
    \item The proposed framework is compared with a recently proposed approach, outperforming it in terms of classification accuracy.
\end{enumerate}

The rest of this paper is structured as follows. Section \ref{section:ViT} presents the proposed ViT model and its working principle. Section 
\ref{section:Results} presents and discusses the performance evaluation results, and carries out a comparative analysis. Finally,  Section \ref{section:Conclusion} concludes this paper.

\begin{figure*}[!ht]
  \centering
  \includegraphics[width=2.00\columnwidth]{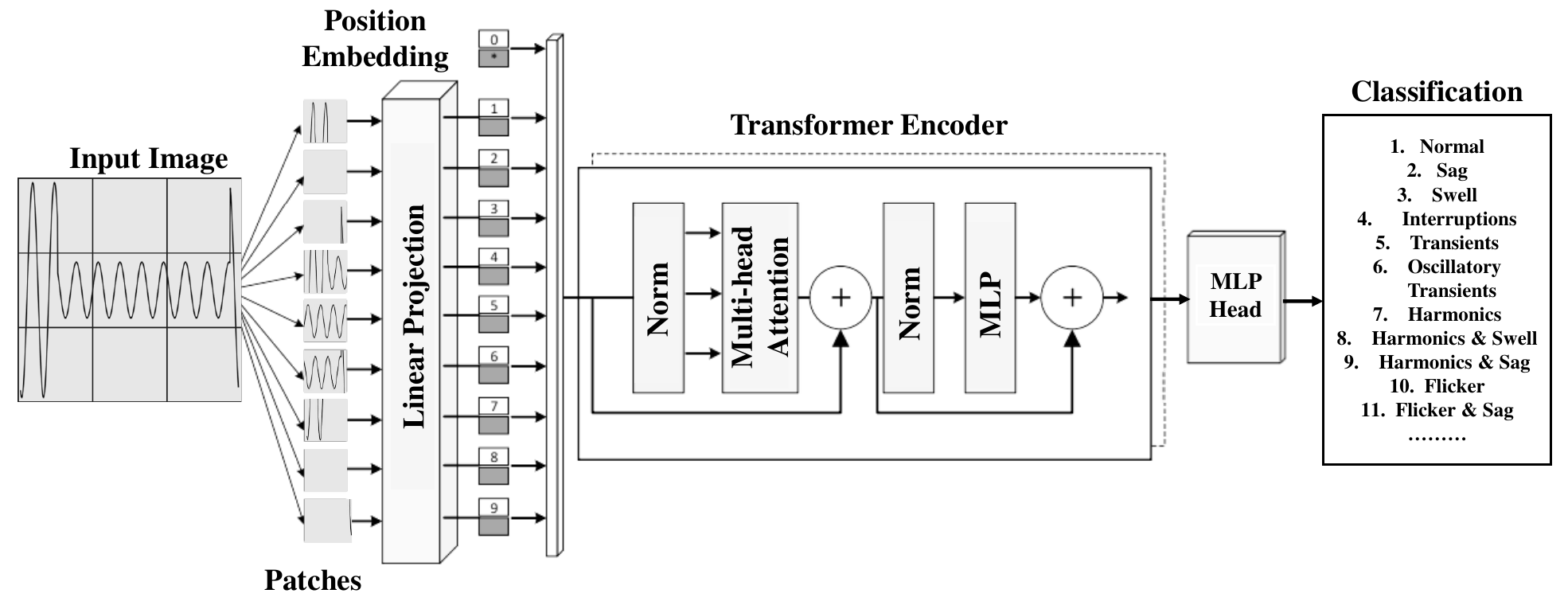}
  \centering
  \caption{Conceptual model for employing ViT for power quality signal's classification. The input image is split into 9 patches for demonstration only. * (0) denotes the extra learnable (class) embedding. }
  \centering
  \label{fig:Conceptual_Model }
\end{figure*}

\section{Vision Transformer for Power Quality Signal Classification }
\label{section:ViT}

Vision Transformer (ViT) leverages the Transformer architecture, initially developed for natural language processing, and applies it to classify PQ signals plotted as images.
In this section, we adapt the ViT model from \cite{DosovitskiyB0WZ21}, whose architecture is shown in Fig. \ref{fig:Conceptual_Model }.
Firstly, each PQ signal is plotted as a 2D image. This image is then divided into a number of same-size patches. These patches are linearly projected by the Vit. Next, the patches' projections and the position/label are embedded by the ViT model into the Transformer Encoder. Afterward, an MLP layer processes the output of the Encoder to determine the inputted signal's class. The details are explained next.

\subsection{Proposed ViT Model Architecture}

\subsubsection{Patch and Position Embedding}

The first layer of the ViT linearly projects the PQ signal image's patches into a lower-dimensional space. Generally, an input PQ signal can be represented as $\mathbf{s} \in \mathbb{R}^{T \times C}$, where $T$ denotes the number of time samples and $C$ denotes the number of channels in the signal.  This PQ signal can be represented as a 2D image. Here,  $\mathbf{x} \in \mathbb{R}^{H \times W \times C}$, where $H$ and $W$ represent the height and width of the  2D representation, respectively \cite{DosovitskiyB0WZ21}. The 2D representation is divided into a grid of non-overlapping patches, each of size $(P, P)$. This results in a total of $N = \frac{H \times W}{P^2}$ patches. Each patch is then flattened into a vector and linearly transformed to a lower-dimensional embedding space using a trainable linear projection:
\begin{equation}
\mathbf{z}_p^0 = [\mathbf{x}_p^1 \mathbf{E}; \mathbf{x}_p^2 \mathbf{E}; \ldots; \mathbf{x}_p^N \mathbf{E}] + \mathbf{E}_{pos},
\end{equation}
where $\mathbf{x}_p^i \in \mathbb{R}^{P^2 \times C}$ is the $i$-th flattened patch, $\mathbf{E} \in \mathbb{R}^{(P^2 \times C) \times D}$ is the trainable linear projection, $\mathbf{E}_{pos} \in \mathbb{R}^{(N+1) \times D}$ represents the position embeddings, and $\mathbf{z}_p^0 \in \mathbb{R}^{(N+1) \times D}$ is the resulting sequence of patch embeddings with an added learnable classification token $\mathbf{z}_0^0$.

\subsubsection{Transformer Encoder}

The patch embeddings are then fed into a standard Transformer encoder, which consists of $L$ layers of Multi-Head Self-Attention (MSA) and Multi-Layer Perceptron (MLP) blocks, with Layer Normalization (LN) applied before each block and residual connections after each block.

\begin{align}
\mathbf{z}_\ell' &= \text{MSA}(\text{LN}(\mathbf{z}_{\ell-1})) + \mathbf{z}_{\ell-1}, \quad \ell = 1, \ldots, L, \\
\mathbf{z}_\ell &= \text{MLP}(\text{LN}(\mathbf{z}_\ell')) + \mathbf{z}_\ell', \quad \ell = 1, \ldots, L,
\end{align}
where $\mathbf{z}_\ell \in \mathbb{R}^{(N+1) \times D}$ is the output of the $\ell$-th layer.

\subsubsection{Classification}

The final hidden state corresponding to the classification token $\mathbf{z}_L^0$ is used as the signal representation, which is passed through a classification head (a multi-layer perceptron) to obtain the final class probabilities:
\begin{equation}
\mathbf{y} = \text{softmax}(\text{MLP}(\mathbf{z}_L^0)).
\end{equation}

\subsection{ViT for Power Quality Signal Classification}

The proposed ViT model can be adapted to classify various types of PQ disturbances through the following steps. Firstly, the PQ signal $\mathbf{s}$ is represented as a 2D image $\mathbf{x}$. Secondly, Patch Embedding os performed. In this step, we divide the 2D representation into patches and linearly embed each patch into a lower-dimensional space. Position embeddings are added to retain the positional information of each patch. Thirdly comes the Transformer Encoding step, where the sequence of embedded patches is processed through the Transformer encoder. The self-attention mechanism allows the model to capture long-range dependencies and interactions between different parts of the signal. Finally, the Classification is performed where the output of the previous step is mapped into one of the predefined classes using an MLP layer. By leveraging the self-attention mechanism, the ViT can effectively capture the complex patterns and interactions in power quality signals, making it a powerful model for PQ disturbance classification in the smart grid.

\section{Performance Evaluation of the Proposed ViT Model}
\label{section:Results}

This section evaluates the performance of the proposed ViT for PQDs classification. The model is trained and tested on a dataset with a wide spectrum of power quality classes, and results are discussed.

\subsection{Dataset Description}

To evaluate the performance of the proposed ViT, a comprehensive power quality classes dataset is utilized \cite{Tian9652053}. Herein, the use of a synthetic data generation approach provides valuable insights in controlled environments, enables the possibility of exploring extreme conditions, and provides detailed information. In this dataset, 17 different power quality disturbances are represented, covering most events that can occur in reality. These classes are depicted in Table \ref{tab:classes}. 
\begin{table*}[t!]
\centering
\begingroup
\caption{Simulated Power Quality Disturbances}
\label{tab:classes}
\begin{tabular}{c| c| c| c| c| c  }\hline
\centering
\makebox{PQ Class  } & \makebox{Description  } 
& \makebox{PQ Class } & \makebox{Description  }
& \makebox{PQ Class  } & \makebox{Description  } \\ \hline
Class 0 & normal (no disturbance) & Class 6 & Harmonics & Class 11 & Flicker \& Swell \\  
Class 1 & Sag & Class 7 & Harmonics \& Sag & Class 12 & Sag \& Oscillatory Transient \\  
Class 2 & Swell & Class 8 & Harmonics \& Swell & Class 13 & Swell \& Oscillatory Transient \\ 
Class 3 & Interruption & Class 9 & Flicker & Class 14 & Sag \& harmonics \\ 
Class 4 & Transients/Spike/Impulse & Class 10 & Flicker \& Sag & Class 15 & Swell \& harmonics \\ 
Class 5 & Oscillatory Transient &   &   & Class 16 & Notch \\ \hline
\end{tabular}
\endgroup
\end{table*}
They cover the normal sine wave (with no disturbance) representing a healthy voltage signal, as well as 16 disturbances. These disturbances can be either: single events, e.g., the interruption, transients, swell and sag events, or multiple simultaneous events, e.g., harmonics with sag and flicker with swell events. More details on the mathematical models used to generate the PQ signals can be found in \cite{Igual8378902,Tian9652053}. Samples are generated at a sampling frequency of 3.2 kHz assuming a 50-Hz fundamental frequency and a unity per-unit amplitude. The length of each power quality signal vector is 650 samples, which is approximately the length of 10 fundamental cycles. To be more realistic, the power quality signals are simulated with an additive white Gaussian noise of 30 dB signal-to-noise ratio, to account for the unavoidable measurement noise \cite{saber2023fast}.

Each class of the 17 is represented by 15,000 samples, making the complete dataset of the size 255,000 samples. Samples are then converted into images. In each signal's image, the value of $H$ $\times$ $W$  is  224 $\times$ 224 pixels. Each image is then divided into 16 patches, resulting in an $N$ of 196. 
A dataset covering the 17 classes is shuffled and randomly split into training and testing datasets following the proportions in \cite{Tian9652053}.

\subsection{Training and Evaluation Metrics}

\begin{figure}[t!]
   \centering  \includegraphics[width=1\columnwidth]{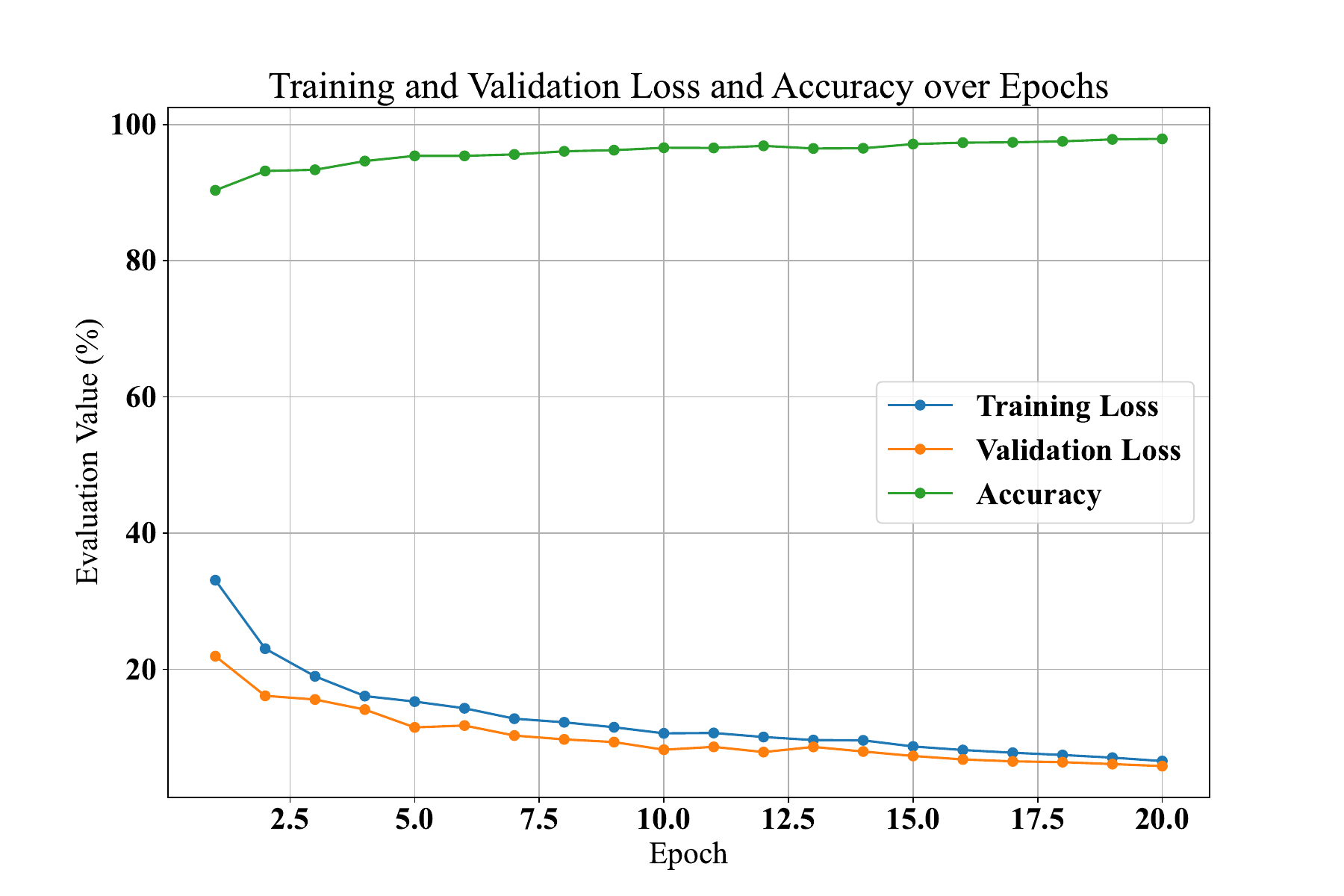}
   \centering
   \caption{Training curve of the ViT model.}
   \centering
  \label{fig:training_curve}
 \end{figure}

A ViT model with the architecture presented in Section \ref{section:ViT} is trained on the training dataset. Throughout this paper, the ViT with the parameters introduced in \cite{DosovitskiyB0WZ21} is leveraged. The number of $C$  equals one for the grey scale PQ signal images in this paper. This ViT is a model that was previously trained on a huge dataset of 14 million images and 21,843 classes, making the model powerful for image classification tasks across different domains.
We adopt the implementation of \cite{Huggingface} for the ViT in this paper.
When the ViT is trained, the model learns the inner representation of signals' images that can then be used to extract features useful for downstream tasks, namely PQD classification in this paper. 
The training is performed for 20 epochs, 
as shown in Fig. \ref{fig:training_curve},
taking around 10 hours. Afterwards, the ViT model is ready for testing. A 64-Bit computer  Core i9 with 128 GB RAM and NVIDIA GeForce RTX 3090 GPU is used  in this experiment. The learning rate was 0.0001, the weight decay was 0.02, the training batch size of 32 and the testing batch size of 8.

In the next step, testing, we evaluate the performance of the ViT using a group of commonly used statistical metrics, which are precision, recall, and f1-score \cite{saber2023cyber}. These metrics are calculated for each class as follows:

\begin{equation}
\text{precision}_i =  \frac{ \text{tp}_i }{ \text{tp}_i + \text{fp}_i }  
\end{equation}

\begin{equation}
\text{recall}_i =  \frac{ \text{tp}_i }{ \text{tp}_i + \text{fn}_i }  
\end{equation}

\begin{equation}
\text{f1-score}_i =  \frac{ 2 \times \text{precision}_i \times \text{recall}_i }{ \text{precision}_i + \text{recall}_i } 
\end{equation}

\noindent where $tp$, $tn$, $fp$, and $fn$ are the true positives, true negatives, false positives, and false negatives, respectively, determined with respect to class $i$.
Moreover, for a holistic evaluation, the average weighted precision, recall, and f1-score values are also obtained for the 17 classes. These are determined as follows:

\begin{equation}
\text{precision}_w = \frac{\sum_{i=1}^{n} w_i \cdot \text{precision}_i}{\sum_{i=1}^{n} w_i}
\end{equation}
 
\begin{equation}
\text{recall}_w = \frac{\sum_{i=1}^{n} w_i \cdot \text{recall}_i}{\sum_{i=1}^{n} w_i}
\end{equation}

\begin{equation}
\text{f1-score}_w = \frac{\sum_{i=1}^{n} w_i \cdot \text{f1-score}_i}{\sum_{i=1}^{n} w_i}
\end{equation}

\noindent where  $n$ is the number of classes, 17 in this paper, and $w_i$, is the weight for each class $i$; here defined as $1/n$.
In addition, we also use the accuracy metric, determined as:

\begin{equation}
\text{accuracy} =  \frac{ \text{tp}  + \text{tn}  }{ \text{tp}  + \text{tn} + \text{fp}  + \text{fn}   }  
\end{equation}

\begin{table*}[!ht]
\centering
\begingroup
\caption{Confusion Matrix for 17 Classes (in Percentages)}
\label{tab:confusion_matrix}
\begin{tabular}{c|ccccccccccccccccc}
\hline
\diagbox[width=11em]{\textit{True Class}}{\textit{Predicted Class}} 
& C0 & C1 & C2 & C3 & C4 & C5 & C6 & C7 & C8 & C9 & C10 & C11 & C12 & C13 & C14 & C15 & C16 \\ \hline
C0 & 99.2 & 0 & 0.2 & 0 & 0 & 0 & 0 & 0 & 0.6 & 0 & 0 & 0 & 0 & 0 & 0 & 0 & 0 \\
C1 & 0 & 99.2 & 0 & 0 & 0 & 0.8 & 0 & 0 & 0 & 0 & 0 & 0 & 0 & 0 & 0 & 0 & 0 \\
C2 & 0 & 0 & 99.4 & 0 & 0 & 0 & 0 & 0 & 0 & 0 & 0 & 0 & 0 & 0 & 0.6 & 0 & 0 \\
C3 & 0 & 0 & 0.2 & 99.6 & 0 & 0 & 0 & 0 & 0 & 0.2 & 0 & 0 & 0 & 0 & 0 & 0 & 0 \\
C4 & 0 & 0 & 0 & 0 & 100 & 0 & 0 & 0 & 0 & 0 & 0 & 0 & 0 & 0 & 0 & 0 & 0 \\
C5 & 0 & 0 & 0 & 0 & 0 & 100 & 0 & 0 & 0 & 0 & 0 & 0 & 0 & 0 & 0 & 0 & 0 \\
C6 & 0 & 0 & 0 & 0 & 0 & 26.2 & 73.8 & 0 & 0 & 0 & 0 & 0 & 0 & 0 & 0 & 0 & 0 \\
C7 & 0 & 0 & 0 & 0 & 0 & 0 & 0 & 100 & 0 & 0 & 0 & 0 & 0 & 0 & 0 & 0 & 0 \\
C8 & 1.0 & 0 & 0 & 0 & 0 & 0 & 0 & 0 & 99.0 & 0 & 0 & 0 & 0 & 0 & 0 & 0 & 0 \\
C9 & 0 & 0 & 0.6 & 0 & 0 & 0 & 0 & 0 & 0 & 99.2 & 0 & 0 & 0 & 0.2 & 0 & 0 & 0 \\
C10 & 0 & 0 & 0 & 0 & 0 & 0 & 0 & 0 & 0 & 0 & 100 & 0 & 0 & 0 & 0 & 0 & 0 \\
C11 & 0 & 0 & 0 & 0 & 0 & 0 & 0 & 0 & 0 & 0 & 0 & 100 & 0 & 0 & 0.2 & 0 & 0 \\
C12 & 0 & 0 & 0 & 0 & 0 & 0 & 0 & 0 & 0 & 0 & 0 & 0 & 100 & 0 & 0 & 0 & 0 \\
C13 & 0 & 0 & 0 & 0 & 0 & 0 & 0 & 0 & 0 & 0 & 0 & 0 & 0 & 99.8 & 0.2 & 0 & 0 \\
C14 & 0 & 0 & 0 & 0 & 0 & 0 & 0 & 0 & 0 & 0 & 0 & 0 & 0 & 1.8 & 98.2 & 0 & 0 \\
C15 & 0 & 0 & 0.8 & 0 & 0 & 0.4 & 0 & 0 & 0 & 0 & 0 & 0.6 & 0 & 0 & 0 & 98.4 & 0 \\
C16 & 0 & 0 & 0 & 0 & 0 & 0 & 0 & 0 & 0 & 0 & 0.2 & 0 & 0 & 0 & 1.2 & 0 & 98.6 \\ \hline
\end{tabular}
\endgroup
\\ \vspace{3pt}Letter "C" stands for "Class", e.g., C0 is Class 0. Classes explanations are given in Table \ref{tab:classes}.
\end{table*}

\begin{table}[t!]
\centering
\begingroup
\caption{Per-Class Performance Results of the Proposed ViT}
\label{tab:performance_results_perclass}
\begin{tabular}{c|c|c|c}\hline
Class & precision & recall & f1-score \\ \hline
Class 0 & 99.00\% & 99.20\% & 99.10\% \\ \hline
Class 1 & 100.00\% & 99.20\% & 99.60\% \\ \hline
Class 2 & 98.22\% & 99.40\% & 98.81\% \\ \hline
Class 3 & 100.00\% & 99.60\% & 99.80\% \\ \hline
Class 4 & 100.00\% & 100.00\% & 100.00\% \\ \hline
Class 5 & 78.49\% & 100.00\% & 87.95\% \\ \hline
Class 6 & 100.00\% & 73.80\% & 84.93\% \\ \hline
Class 7 & 100.00\% & 100.00\% & 100.00\% \\ \hline
Class 8 & 99.40\% & 99.00\% & 99.20\% \\ \hline
Class 9 & 99.80\% & 99.20\% & 99.50\% \\ \hline
Class 10 & 100.00\% & 100.00\% & 100.00\% \\ \hline
Class 11 & 99.80\% & 99.80\%  & 99.80\% \\ \hline
Class 12 & 99.40\% & 100.00\% & 99.70\% \\ \hline
Class 13 & 98.04\% & 99.80\% & 98.91\% \\ \hline
Class 14 & 99.80\% & 98.20\% & 98.99\% \\ \hline
Class 15 & 98.01\% & 98.20\% & 98.11\% \\ \hline
Class 16 & 100.00\% & 98.60\% & 99.30\% \\ \hline
\end{tabular}
\endgroup
\end{table}

\subsection{Simulation Results}

\begin{figure}[t!]
  \centering
  \includegraphics[width=0.49\columnwidth]{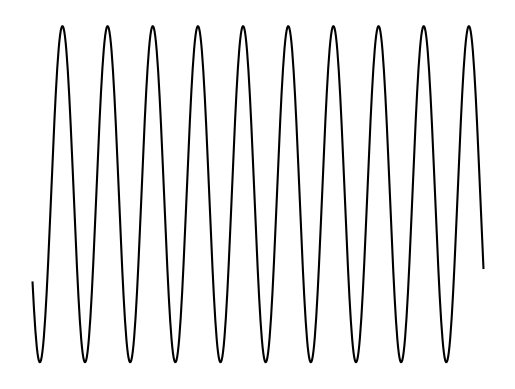} 
  \includegraphics[width=0.49\columnwidth]{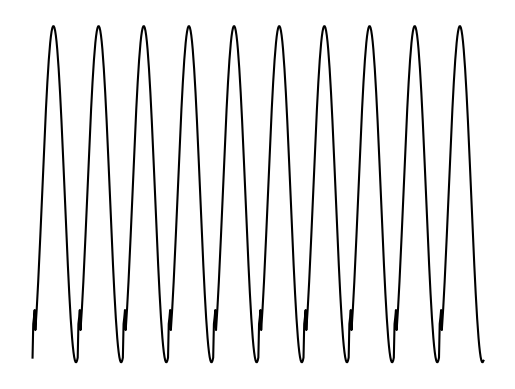} 
  \\
    (a) \hspace{40mm} (b) \\
  \includegraphics[width=0.49\columnwidth]{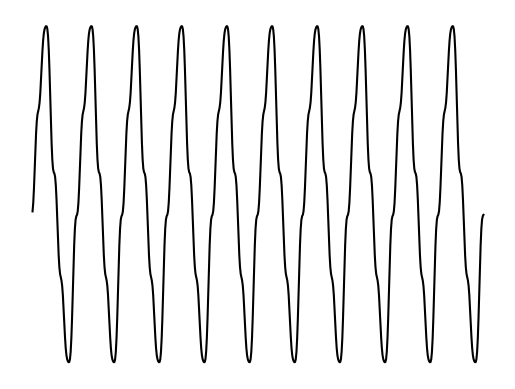} 
    \includegraphics[width=0.49\columnwidth]{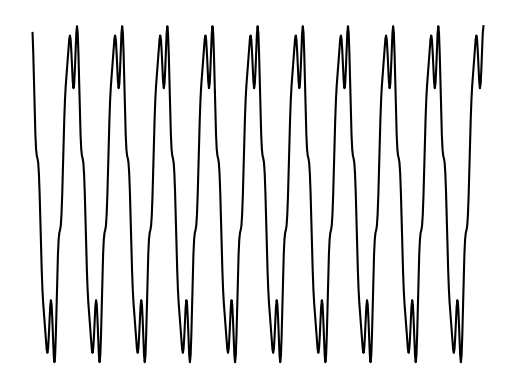}\\
    (c) \hspace{40mm} (d) \\
      \includegraphics[width=0.49\columnwidth]{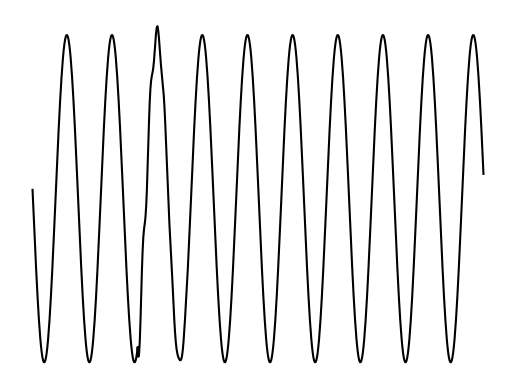}
        \includegraphics[width=0.49\columnwidth]{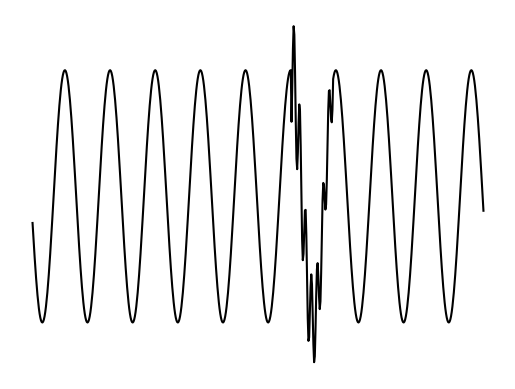} \\
        (e) \hspace{40mm} (f)
  \centering
  \caption{Samples of the PQD Signals, 
  (a) pure sine wave,
  (b) Notch,
  (c) harmonics with small amplitudes, 
  (d) harmonics with large amplitudes, 
  (e) an oscillatory transient with a small amplitude, 
  (f) an oscillatory transient with a large amplitude}
  \centering
  \label{fig:PQDsamples}
\end{figure}

The model is then tested on the initially held-out previously unseen testing dataset. Our results, summarized in Table \ref{tab:perfromance_results_summary}, confirm that the proposed ViT can accurately classify power quality disturbances. 
Fig.
  \ref{fig:PQDsamples}
shows samples from the PQD dataset. 
In detail, most PQD classes are correctly classified with precision and recall of more than 98\%, as depicted in Table \ref{tab:performance_results_perclass}.
Nevertheless, in some cases, the ViT model confuses Class 6 (Harmonics)  with Class 5 (Oscillatory Transients).  This confusion could be mainly because these two classes are similar when plotted as images when the amplitudes of the respective disturbance are small  (e.g. when the amplitude of the transient event is minimal which could look like a harmonic event with short-duration high-frequency harmonics of minimal amplitudes). This point requires further investigation for completeness. 
Aside from this point, the ViT model accurately classifies 97.88 \% of the 17 PQD classes with a precision of 98.23\%, recall of 97.88\%, and f1-score of 97.86\%, demonstrating the ViT model's effectiveness.  
The detailed confusion matrix of testing the proposed ViT model is depicted in Table \ref{tab:comparison}.

\subsection{Performance Comparison}

Further, this section compares the performance of the proposed ViT with that of the recently proposed method. In \cite{Tian9652053}, a CNN-based framework was recently presented for power quality disturbances classification.
The CNN model consists of 3 main units, followed by a dense layer, a batch normalization layer, another dense layer, and finally the classification layer. Each main unit consists of two successive convolutional layers, followed by a max-pooling layer and then a batch normalization layer. 
The CNN was applied to the same dataset used in this paper.   
The comparative results shown in Table \ref{tab:comparison} show that the proposed ViT model outperforms the CNN model in terms of PQ classification testing accuracy.

\begin{table}[t!]
\centering
\begingroup
\caption{Summary of Proposed ViT's Performance Results}
\label{tab:perfromance_results_summary}
\begin{tabular}{c| c| c| c  }\hline
\centering
\makebox{$accuracy$} & \makebox{precision} & \makebox{recall} & \makebox{f1-score} \\ \hline
97.88\% &  98.23\%  &  97.88\% &  97.86\% \\  \hline
\end{tabular}
\endgroup
\end{table}

\begin{table}[!ht]
\centering
\begingroup
\caption{Comparative Analysis: Comparison of Testing Accuracies
}
\label{tab:comparison}
\begin{tabular}{c| c   }\hline
\centering
\makebox{Approach} & \makebox{Classification $accuracy$} \\ \hline
 Proposed ViT & 
97.88\%
 \\  \hline
CNNs \cite{Tian9652053}  & 92.01\% \\  \hline
\end{tabular}
\endgroup
\end{table}

\section{Conclusion}
\label{section:Conclusion}

This paper presented a novel framework for accurate power quality events classification using ViTs, which have not been applied before by any of the related works. The proposed ViT achieves high classification accuracy by learning the inner representations in PQ signals, represented as images, and captures the global dependencies in the image through self-attention mechanisms. The proposed framework was trained and tested on a comprehensive and diverse dataset comprising 17 different classes of PQ signals, and more than 250,000 samples. Our experimental results confirmed that the proposed ViT-based approach accurately classifies PQ disturbances, even under measurement noise, outperforming recently proposed techniques.

\bibliography{Manuscript.bib}

\begin{thebibliography}{10}
\providecommand{\url}[1]{#1}
\csname url@samestyle\endcsname
\providecommand{\newblock}{\relax}
\providecommand{\bibinfo}[2]{#2}
\providecommand{\BIBentrySTDinterwordspacing}{\spaceskip=0pt\relax}
\providecommand{\BIBentryALTinterwordstretchfactor}{4}
\providecommand{\BIBentryALTinterwordspacing}{\spaceskip=\fontdimen2\font plus
\BIBentryALTinterwordstretchfactor\fontdimen3\font minus \fontdimen4\font\relax}
\providecommand{\BIBforeignlanguage}[2]{{%
\expandafter\ifx\csname l@#1\endcsname\relax
\typeout{** WARNING: IEEEtran.bst: No hyphenation pattern has been}%
\typeout{** loaded for the language `#1'. Using the pattern for}%
\typeout{** the default language instead.}%
\else
\language=\csname l@#1\endcsname
\fi
#2}}
\providecommand{\BIBdecl}{\relax}
\BIBdecl

\bibitem{Tian9652053}
J.~Tian, B.~Wang, J.~Li, and Z.~Wang, ``Adversarial attacks and defense for cnn based power quality recognition in smart grid,'' \emph{IEEE Transactions on Network Science and Engineering}, vol.~9, no.~2, pp. 807--819, 2022.

\bibitem{singh2009power}
G.~K. Singh, ``Power system harmonics research: a survey,'' \emph{European Transactions on Electrical Power}, vol.~19, no.~2, pp. 151--172, 2009.

\bibitem{chawda2020comprehensive}
G.~S. Chawda, A.~G. Shaik, M.~Shaik, S.~Padmanaban, J.~B. Holm-Nielsen, O.~P. Mahela, and P.~Kaliannan, ``Comprehensive review on detection and classification of power quality disturbances in utility grid with renewable energy penetration,'' \emph{IEEE Access}, vol.~8, pp. 146\,807--146\,830, 2020.

\bibitem{saber2024enhancing}
A.~M. Saber, A.~Youssef, D.~Svetinovic, H.~Zeineldin, D.~Kundur, and E.~El-Saadany, ``Enhancing power quality event classification with ai transformer models,'' in \emph{2024 IEEE Power \& Energy Society General Meeting (PESGM)}, 2024, pp. 1--6.

\bibitem{ahila2015integrated}
R.~Ahila, V.~Sadasivam, and K.~Manimala, ``An integrated pso for parameter determination and feature selection of elm and its application in classification of power system disturbances,'' \emph{Applied Soft Computing}, vol.~32, pp. 23--37, 2015.

\bibitem{khokhar2017new}
S.~Khokhar, A.~A.~M. Zin, A.~P. Memon, and A.~S. Mokhtar, ``A new optimal feature selection algorithm for classification of power quality disturbances using discrete wavelet transform and probabilistic neural network,'' \emph{Measurement}, vol.~95, pp. 246--259, 2017.

\bibitem{wang2019novel}
S.~Wang and H.~Chen, ``A novel deep learning method for the classification of power quality disturbances using deep convolutional neural network,'' \emph{Applied energy}, vol. 235, pp. 1126--1140, 2019.

\bibitem{zhao2019power}
W.~Zhao, L.~Shang, and J.~Sun, ``Power quality disturbance classification based on time-frequency domain multi-feature and decision tree,'' \emph{Protection and Control of Modern Power Systems}, vol.~4, pp. 1--6, 2019.

\bibitem{liu2020high}
J.~Liu, H.~Song, H.~Sun, and H.~Zhao, ``High-precision identification of power quality disturbances under strong noise environment based on fastica and random forest,'' \emph{IEEE Transactions on Industrial Informatics}, vol.~17, no.~1, pp. 377--387, 2020.

\bibitem{rodriguez2021classification}
M.~A. Rodriguez, J.~F. Sotomonte, J.~Cifuentes, and M.~Bueno-L{\'o}pez, ``A classification method for power-quality disturbances using hilbert--huang transform and lstm recurrent neural networks,'' \emph{Journal of Electrical Engineering \& Technology}, vol.~16, pp. 249--266, 2021.

\bibitem{ma2023complex}
J.~Ma, Q.~Tang, M.~He, L.~Peretto, and Z.~Teng, ``Complex pqd classification using time--frequency analysis and multiscale parallel attention residual network,'' \emph{IEEE Transactions on Industrial Electronics}, 2023.

\bibitem{cui2022detection}
C.~Cui, Y.~Duan, H.~Hu, L.~Wang, and Q.~Liu, ``Detection and classification of multiple power quality disturbances using stockwell transform and deep learning,'' \emph{IEEE Transactions on Instrumentation and Measurement}, vol.~71, pp. 1--12, 2022.

\bibitem{sindi2021novel}
H.~Sindi, M.~Nour, M.~Rawa, {\c{S}}.~{\"O}zt{\"u}rk, and K.~Polat, ``A novel hybrid deep learning approach including combination of 1d power signals and 2d signal images for power quality disturbance classification,'' \emph{Expert Systems with Applications}, vol. 174, p. 114785, 2021.

\bibitem{eristi2022new}
B.~Eristi and H.~Eristi, ``A new deep learning method for the classification of power quality disturbances in hybrid power system,'' \emph{Electrical Engineering}, vol. 104, no.~6, pp. 3753--3768, 2022.

\bibitem{ozer2021cnn}
{\.I}.~{\"O}zer, S.~B. Efe, and H.~{\"O}zbay, ``Cnn/bi-lstm-based deep learning algorithm for classification of power quality disturbances by using spectrogram images,'' \emph{International Transactions on Electrical Energy Systems}, vol.~31, no.~12, p. e13204, 2021.

\bibitem{9832923}
D.~L. Carní and F.~Lamonaca, ``Toward an automatic power quality measurement system: An effective classifier of power signal alterations,'' \emph{IEEE Transactions on Instrumentation and Measurement}, vol.~71, pp. 1--8, 2022.

\bibitem{DosovitskiyB0WZ21}
\BIBentryALTinterwordspacing
A.~Dosovitskiy, L.~Beyer, A.~Kolesnikov, D.~Weissenborn, X.~Zhai, T.~Unterthiner, M.~Dehghani, M.~Minderer, G.~Heigold, S.~Gelly, J.~Uszkoreit, and N.~Houlsby, ``An image is worth 16x16 words: Transformers for image recognition at scale,'' in \emph{9th International Conference on Learning Representations, {ICLR} 2021, Virtual Event, Austria, May 3-7, 2021}.\hskip 1em plus 0.5em minus 0.4em\relax OpenReview.net, 2021. [Online]. Available: \url{https://openreview.net/forum?id=YicbFdNTTy}
\BIBentrySTDinterwordspacing

\bibitem{Igual8378902}
R.~Igual, C.~Medrano, F.~J. Arcega, and G.~Mantescu, ``Integral mathematical model of power quality disturbances,'' in \emph{2018 18th International Conference on Harmonics and Quality of Power (ICHQP)}, 2018, pp. 1--6.

\bibitem{saber2023fast}
A.~M. Saber, A.~Selim, V.~Kadkikar, H.~Zeineldin, and E.~El-Saadany, ``Fast deep-learning-based recognition of multiple power quality events under noise and dc offset,'' in \emph{2023 IEEE Conference on Power Electronics and Renewable Energy (CPERE)}.\hskip 1em plus 0.5em minus 0.4em\relax IEEE, 2023, pp. 1--6.

\bibitem{Huggingface}
``Vit-base-patch16-224-in21k,'' Available: \url{https://huggingface.co/google/vit-base-patch16-224-in21k}, accessed: 1 April 2024.

\bibitem{saber2023cyber}
A.~M. Saber, A.~Youssef, D.~Svetinovic, H.~H. Zeineldin, and E.~F. El-Saadany, ``Cyber-immune line current differential relays,'' \emph{IEEE Transactions on Industrial Informatics}, 2023.

\end{thebibliography}
\bibliographystyle{IEEEtran}
\end{document}